\documentclass[fleqn,twoside]{article}
\usepackage{espcrc2}
\newcommand{\be}{\begin{equation}}
\newcommand{\ee}{\end{equation}}
\newcommand{\bea}{\begin{eqnarray}}
\newcommand{\eea}{\end{eqnarray}}

\newcommand{\psl}{\hbox{$\partial$\hspace{-.5em}/}}
\begin{document}
\hyphenation{author another created financial paper re-commend-ed Post-Script}
\title{Lepton Flavour Violation in the Minimal $SO(10)$ GUT Model and 
in the Standard Model with additional Heavy Dirac Neutrinos}
\author{Takeshi Fukuyama
\footnote{E-Mail: fukuyama@se.ritsumei.ac.jp}
\address[RU]{Department of Physics, 
Ritsumeikan University, Kusatsu, Shiga, 525-8577 Japan}, 
Amon Ilakovac
\footnote{E-Mail: ailakov@rosalind.phy.hr}
\address[UZ]{University of Zagreb, 
Department of Physics, Bijeni\v cka cesta 32, 
HR-10002 Zagreb, Croatia}, 
Tatsuru Kikuchi
\footnote{E-Mail: rp009979@se.ritsumei.ac.jp}
\addressmark[RU]
,
 Stjepan Meljanac\address[RB]
   {Institut Rudjer Bo\v skovi\'c, Bijeni\v cka cesta 54, 
HR-10002 Zagreb, Croatia}}
\begin{abstract}

Lepton flavour violation is considered in two models, 
Standard Model with additional heavy Dirac neutrinos 
and minimal supersymmetric standard model (MSSM) with 
$SO(10)$ theoretical frame and minimal SUGRA (mSUGRA) 
SUSY-breaking mediation mechanism. The models are briefly 
explained. The structure of the amplitudes for neutrinoless 
decays of the lepton into a lepton and semileptonic neutrinoless 
decays of a lepton is given. The method for identification of 
meson content in the quark currents is exposed. The comparison 
of the amplitudes is made showing which of the 
processes have the largest branching ratios. 
The numerical results for the dominant decay rates are presented.

\vspace{1pc}
\end{abstract}

\maketitle

\section{Introduction}

Neutrino oscillation data not only show that the neutrinos 
do have mass and do mix, but these mixings and differences 
of the squares of the neutrino masses are well determined. 
Extensions of the Standard Model (SM) are therefore necessary, 
and any such extension must be able to describe the neutrino 
data together with all other experimental data which SM describes well. 
Two of such models are considered here. The simpler of these models, 
SM extended by heavy (quasi-)Dirac neutrinos \cite{EW86,RMiV86,Val90} 
(called model A here) has only additional heavy (quasi-) Dirac neutrino 
fields compared to the SM field content. 
The second model \cite{Babu93,FO02,Goh03,Senj03,Mim04,Mat04} (model B) 
is based on the minimal supersymmetric standard model (MSSM) with 
heavy right-handed neutrinos \cite{Masiero86,Hisano96}, 
with model parameters determined by the underlying GUT model and through 
the minimal SUGRA (mSUGRA) SUSY-breaking mediation mechanism \cite{mSUGRA}. 

The existence of the neutrino mixings implies that the lepton flavours 
are violated. An independent confirmation of the lepton flavour violation 
(LFV) is necessary. The last improvements of the 
LFV upper bounds \cite{YE_tau04} strongly motivate further theoretical 
investigation of these processes. The structure of the LFV amplitudes 
and a numerical study of the corresponding decay rates in model A and 
model B are the subject of of this paper. The main topic of this paper 
are semileptonic (SL) neutrinoless LFV decays of charged leptons. 
The LFV decays of the charged lepton into lepton and photon 
($\ell \to \ell'\gamma$) are considered also, to determine the bounds 
on the parameters in the model B. 

\section{Models}

The model A is inspired by a $E_8 \times E_8$ heterotic-
superstring theory. Along with the SM neutrino fields, 
it contains a number of right-handed neutrino fields and 
equal number of left-handed $E_6$-singlet neutrino fields 
\cite{EW86,RMiV86,Val86}. The right-handed neutrino fields 
mix through the Dirac-type Yukawa couplings with SM neutrino fields 
and right-handed neutrino fields. The mass matrix has three zero 
eigenvalues while the other eigenvalues are are large (typically 
$\sim 1\ {\rm GeV}$). The masses and the mixings of the light neutrinos 
may be induced through the small Majorana-type Yukawa couplings of 
the left-handed $E_6$ singlet neutrino fields \cite{Val86,GGVal89}. 
Fitting of the neutrino oscillation data is easily achievable. 
The small Majorana mass terms promote the heavy Dirac fields to 
heavy quasi-Dirac fields. The structure of the mass matrix without 
Majorana terms assures conservation of the total lepton number 
\cite{Val86,BerVal87,RiVal90}, but the individual lepton flavours are 
violated through the neutrino mixings 
\cite{BerVal87,RiVal90,GGVal92,AmApos95}. Through the neutrino mixings, 
the SM neutrino fields become massless (light) neutrino fields 
($\nu_i$, $i=1,2,3$) with small admixture of the heavy (quasi-)Dirac 
fields ($N_i$, $i=4,\cdots$). Through the loops, in which 
only the massive neutrinos give a contribution, the heavy 
(quasi-)Dirac neutrino components of the SM neutrino field give 
rise to the LFV effects that are not GIM suppressed. 
The CKM-type matrix ($B_{l_in_j}$) which contains neutrino mixing 
matrix is not known in details. Only the upper limits, 
$(s_L^{\nu_i})^2=\sum_{j=4,\cdots} |B_{l_iN_j}|^2$ 
\cite{sL88_92,sLTomass95}, on the combinations of its elements 
are limited experimentally \cite{Tomass94}. 
The leading-order LFV contributions come from the loops with two 
or four $B_{lN}$ matrices (there is one tree-level contribution, 
but it is negligible compared to the loop contributions). 
The combinations of matrix elements which appear in the loop 
contributions to the LFV amplitudes cannot be determined, too. 
Therefore, for given $(s_L^{\nu_i})^2$-s only the upper limits of the 
LFV amplitudes can be found. 

In model B the main sources of LFV are lowest-order loop amplitudes 
with two or four lepton-slepton-chargino/neutralino vertices \cite{Hisano96}. 
The constants in these vertices are defined by mSUGRA SUSY breaking mediation 
mechanism \cite{mSUGRA} and the underlying GUT model, here chosen to be 
the minimal $SO(10)$ model \cite{Babu93,FO02,Goh03,Senj03,Mim04,Mat04}. 
The model contains two $SO(10)$ Higgs superfields, ${\bf 10}$ and 
${\bf \overline{126}}$ in the Yukawa sector, and additional Higgs 
superfields ${\bf 126}$ and ${\bf 210}$ to preserve the supersymmetry 
at the GUT scale \cite{TATSNI04} and to break the $SO(10)$ symmetry to 
$SU(3)_c \times SU(2)_L\times U(1)$ ($G_{321}$) symmetry 
\cite{TATSNI04,TATSNII04}. By assumptions of the mSUGRA mediation 
mechanism, the soft SUSY breaking terms at the GUT scale 
(the mass term and the trilinear interaction term) satisfy the 
universality conditions, according to which the mass soft SUSY-breaking
term is flavour-diagonal. In the renormalization group (RG) 
evolution from the GUT scale to the SM scale, mass soft SUSY-breaking 
terms receive the off-diagonal LFV contributions through the Dirac-neutrino 
Yukawa couplings \cite{Masiero86}. Diagonalization of the sfermion and 
gaugino mass matrices leads to the chargino and neutralino mass eigenstates 
with LFV lepton-slepton-chargino/neutralino vertices. 
Only three parameters, defining the universal soft SUSY-breaking parameters 
at the GUT scale, have to be introduced in this procedure. Concerning the 
mixing matrices in model B, which define the mass eigenstates, 
they are known in details. Therefore, for a given set of initial parameters, 
LFV amplitudes can be precisely evaluated. That allows to determine 
both the lower and upper bounds of LFV decay rates \cite{TAT04}. 

Concerning the RGE procedure there are two subtleties. First, as there 
are only two Higgs superfields in the Yukawa superpotential, the lepton 
and quark masses at the GUT scale are defined in terms of two mass matrices. 
That leads to two equations relating the quark and lepton masses at the GUT 
scale, which depend on two complex parameters \cite{FO02}. The quark and 
lepton masses at the GUT scale are obtained RG-evoluting them from the 
experimental values at the scale of the $Z$-boson mass. Inserting them into 
the two equations one can find the solutions for the two complex parameters 
(the phase of one parameter cannot be determined). With that, fermion mass 
matrices (and corresponding Yukawa matrices) are determined at the GUT scale, 
except for the right-neutrino mass matrix. To determine the right-handed neutrino 
matrix, one more complex parameter has to be introduced \cite{FO02}. 
With right-neutrino
masses determined, the light neutrino masses can be found using the type I 
see-saw mechanism. RG-evoluting the light neutrino operator 
one obtains the light-neutrino mass matrix at the low-energy scale 
which has to agree with the present neutrino-oscillation data. 
The agreement can be achieved. In fact, fitting of the neutrino-oscillation 
data is taken as a necessary condition that has to be fulfilled before 
proceeding to further calculations. That means that in the LFV breaking 
couplings, the information on the experimental neutrino oscillation data 
is built in. Second, the $SO(10)$ model predicts the proton decay. 
The fitting of the model parameters can be achieved to satisfy the 
present proton decay lower bound \cite{TATSNIII04}. 
A detailed analysis of the Higgs mass matrices 
is needed to obtain the parameters defining the proton decay, and RG 
evolution of these parameters to the proton mass scale is necessary to obtain 
the the effective proton-decay Lagrangian \cite{TATSNI04,TATSNII04}. 
Therefore, in principle, model B can relate branching ratios for 
neutrinoless LFV decays of charged leptons with the neutrino-oscillation 
data and lifetime of the proton.

\section{Effective Lagrangians for LFV interactions}

In any model containing SM as its low-energy limit, the following effective 
Lagrangians describe the neutrinoless LFV decays
\bea
\label{Lga}
{\cal L}^{\mathrm{eff}}_{\ell_i \ell_j \gamma} &=&
-e\: \overline{\ell}_j
[(-\partial^2 \gamma_\mu + \psl \partial_\mu) A^\mu
\nonumber\\&&
\times
  ({\cal P}^L_{1\gamma} P_L + {\cal P}^R_{1\gamma} P_R)
\nonumber\\&&
+
\sigma_{\mu\nu}\partial^\nu A^\mu
({\cal P}^L_{2\gamma} P_L + {\cal P}^R_{2\gamma} P_R)] 
\ell_i,
\\
\label{LZ}
{\cal L}^{\mathrm{eff}}_{\ell_i \ell_j Z} &=&
g \overline{\ell}_j
[\gamma_\mu Z^\mu ({\cal P}^L_{Z} P_L + {\cal P}^R_{Z} P_R)]
\ell_i,
\\
\label{LH}
{\cal L}^{\mathrm{eff}}_{\ell_i \ell_j H} &=&
g \overline{\ell}_j  
[{\cal P}^L_{H} P_L + {\cal P}^R_{H} P_R] \ell_i H,
\\
\label{Lbox}
{\cal L}^{\mathrm{box}}_{\mathrm{eff}} &=&
\sum_{\bar{q}_aq_b} \sum_{X X'} \sum_{\Gamma}
  {\cal B}^{X X'}_{\Gamma ab},
\nonumber\\&&
\times
(\bar{\ell}_j \Gamma P_X \ell_i)
(\bar{q}_a \Gamma P_{X'} q_b)
\eea
where $P_{L,R}=(1\pm\gamma_5)/2$. The dummy variables of the sums 
in the last expression assume following values 
$\bar{q}_aq_b = \bar{u}u, \bar{d}d, \bar{s}s, \bar{d}s, \bar{s}d$, 
$X,X' = LL,RR,LR$, $\Gamma = \gamma_\mu,1,\sigma_{\mu\nu}$. 
The Lagrangians contain the SM fields only. The effective Lagrangians 
describe the decays $\ell \to \ell'\gamma^*$, 
$\ell \to \ell' Z^*$, $\ell \to \ell' H^*$ and $\ell \to \ell'q_a q_b$, 
respectively, where $\ell$ and $\ell'$, $\gamma^*$, $Z^*$ and $q_a$ 
and $q_b$ denote leptons, photon, Z-boson, Higgs and quarks, respectively 
(star denotes an off-mass-shell particle). 
The effective Lagrangian for the decay of lepton into three leptons 
is not given because we do not consider these decays. They have been 
considered previously in model A \cite{GGVal92,AmApos95} 
and B \cite{Hisano96}. It has almost the same Lorentz structure as 
the effective Lagrangian for $\ell \to \ell' q_a q_b$. 
The information on LFV induced by a model induces is contained in the form factors 
${\cal P}^L_{1,2\gamma}$, ${\cal P}^R_{1,2\gamma}$, 
${\cal P}^L_{Z}$, ${\cal P}^R_{Z}$, ${\cal P}^L_{H}$, 
${\cal P}^R_{H}$, ${\cal B}^{X X'}_{\Gamma ab}$. 
Form factors contain the loop functions of loop diagrams describing 
the LFV processes. The structure of the "photon" effective Lagrangian 
reflects the gauge invariance of the corresponding amplitude. 
The effective Lagrangian for the "box" process $\ell \to \ell' q_a q_b$ 
is the most general one. In many models just few of the form factors 
are different from zero.  Model A contains only terms bilinear in 
$\gamma_\mu P_{L,R}$ matrices, while model B contains all form factors. 

Using the effective Lagrangians the neutrinoless SL LFV, amplitudes of 
charged leptons and amplitudes for $\ell \to \ell' \gamma$ processes 
may be found at the quark-lepton level. 

\section{Hadronization procedure}

To obtain the amplitudes of neutrinoless SL LFV charged lepton decays 
in terms of lepton and meson fields, the quark currents have to be 
transformed into the corresponding meson fields. Here only the decays with 
one or two pseudoscalar mesons or one vector meson are considered. 
The conversion of the axial-vector and vector currents 
into pseudoscalar and vector mesons, respectively is achieved invoking 
the PCAC hypothesis \cite{Marshak69} and VMD (vector-meson dominance) 
hypothesis \cite{Sakurai69,Fubini73,Zielinski87}, respectively. 
The scalar and pseudoscalar currents are hadronized comparing the QCD 
quark Lagrangian with the corresponding effective meson Lagrangian 
\cite{Gerard87}. The tensor currents are hadronized using the equations 
of motion for current quarks \cite{GassLeut84} and VMD hypothesis. 
The semileptonic LFV amplitudes for processes with two pseudoscalar 
mesons in the final state have scalar-quark current contribution and 
vector-quark current contributions through the vector-meson resonances 
decaying into two pseudoscalar mesons. For that purpose the Lagrangian 
of vector-meson---pseudoscalar-meson interactions is introduced 
\cite{Bando88,Bando85}.

\section{Comparison of the branching ratios}

With amplitudes at lepton-meson level, the comparison and evaluation of the 
decay rates of the neutrinoless SL LFV charged lepton decays and 
$\ell \to \ell' \gamma$ decays can be performed. For comparison of 
decay rates it is useful to classify them in terms of the basic subamplitudes 
they contain, ${\cal A}_\gamma$, ${\cal A}_Z$, ${\cal A}_H$ and 
${\cal A}_{b_{lq}}$, which correspond to the effective Lagrangians 
(\ref{Lga}), (\ref{LZ}), (\ref{LH}) and (\ref{Lbox}), respectively. 
Such a classification is given in Table 1 for the processes considered 
here. \\[.15cm] 

\noindent
\begin{minipage}[t]{7cm}
\noindent
Table 1\\
{\small The decay rates for the charged lepton neutrinoless 
SL LFV processes and $\ell \to \ell'\gamma$ processes.}
{\small
\begin{tabular}{lll}\\
\hline\hline
process  & model A & model B \\
\hline
$\ell \to \ell' \gamma$  & $\gamma$ & $\gamma$\\
$(\tau\to \ell P^0)_c$  &
   $\gamma$, $Z$, $b_{lq}$ &
   $\gamma$, $Z$, $b_{lq}$ \\
$(\tau\to \ell P^0)_n$  &
   $\ell$-$q$-$box$ & $b_{lq}$\\
$(\tau\to \ell V^0)_c$  &
   $\gamma$, $Z$, $b_{lq}$ & $\gamma$, $Z$, $b_{lq}$\\
$(\tau\to \ell V^0)_n$ &
   $b_{lq}$ & $b_{lq}$\\
$(\tau^-\to \ell'^-P_1P_2)_c$ &
   $\gamma$, $Z$, $H$, $b_{lq}$ &
   $\gamma$, $Z$, $H$, $b_{lq}$ \\
$(\tau^-\to \ell'^-P_1P_2)_n$ &
   $b_{lq}$, $W^+W^-$ & $b_{lq}$ \\
$(\tau^-\to \ell'^-P_1P_2)_{c,H}$ &
   $H$, $W^+W^-$ & $H$ \\
\hline\hline
\end{tabular}
}\\[2pt]
\footnotesize{
c(n) denotes (non)conservation of the quark flavour}
\end{minipage}\\[.15cm]

In model A there is a window of the heavy neutrino masses ($m_N$) where the 
matrix elements $B_{lN}$ are not constrained by heavy neutrino mass ($m_N$) 
values. Within that window, the functional dependence of the subamplitudes 
corresponds to the functional dependence of loop functions constituting the 
subamplitudes. Speciffically, the  subamplitudes ${\cal A}_\gamma$, 
${\cal A}_Z$, ${\cal A}_H$ and ${\cal A}_{b_{lq}}$ behave as 
$\ln m_N$, $m_N^2$, $m_N^2$ and $\ln m_N$, respectively. 
The ${\cal A}_Z$ has the largest combination of couplings and if it 
contributes to a LFV amplitude, it determines its behaviour in the large mass 
limit. For very large masses, perturbative unitarity bound forces $B_{lN}$ 
to behave as inverse of the heavy neutrino mass leading to $\ln m_N/m_N^2$, 
$1/m_N^2$, $1/m_N^2$ and $\ln m_N/m_N^4$ behaviour of the amplitudes 
${\cal A}_\gamma$, ${\cal A}_Z$, ${\cal A}_H$ and ${\cal A}_{b_{lq}}$ 
respectively. That behaviour assures that there is no LFV when the heavy 
neutrino mass tends to infinity. Comparing the expressions for the branching 
ratios in which only the dominant subamplitude is retained, one finds that 
there are only two groups of processes that are experimentally interesting, 
one with ${\cal A}_\gamma$ only ($\ell \to \ell'\gamma$ processes), 
and the other with the ${\cal A}_Z$ amplitude (all quark-flavour conserving 
processes. 
The processes with box amplitude are strongly suppressed by their $m_N$ 
dependence and couplings in the vertices. The Higgs amplitude is strongly 
suppresses by the Higgs Yukawa couplings. Within the group with 
${\cal A}_\gamma$ amplitude, by far the best process is $\mu \to e\gamma$. 
The branching ratios for the processes with the ${\cal A}_Z$ amplitude 
are all of the same order of magnitude, the $\tau\to\pi^0$, and 
$\tau\to\rho$ being somewhat better than the other processes. 

In model B, the characteristic scale of new physics is the mass of the 
lightest SUSY particle $m_{\rm SUSY}$. Not only the loop functions 
in the ${\cal A}_\gamma$, ${\cal A}_Z$ and ${\cal A}_{b_{lq}}$ have the 
softer behaviour than in model A, $1/m^2_{\rm SUSY}$, $\ln m_{\rm SUSY}$ 
and $1/m^2_{\rm SUSY}$ respectively, but all coupling constants decrease 
with $m_{\rm SUSY}$ at least as fast as $1/m_{\rm SUSY}$. 
That leads to the dominance of the ${\cal A}_\gamma$ 
(photon penguin amplitude) in the large mass limit. 
Again, there is no LFV in the limit when the characteristic mass of the new 
physics tends to infinity. All branching ratios decrease as $m_{\rm SUSY}$ 
is increased, opposite to the behaviour of branching ratios in model A. 
We do not discuss the SM Higgs amplitude, because the model B has five 
Higgs bosons and some of them may have the LFV at the tree level. 
Limit on the tree-level Higgs contribution constraints combination of 
the Higgs boson mass and its couplings. In model B, the region of allowed 
values of mass parameter $m_{\rm SUSY}$ is small. As explained above, the couplings 
are precisely known when the parameters of the model are given. Therefore, 
upper and lower bound for $m_{\rm SUSY}$ can be determined if there are 
adequate experimental data. The lower $m_{\rm SUSY}$ bound 
($m_{\rm SUSY}=560\ {\rm GeV}$) is obtained from the $\mu\to e\gamma$ process. 
The upper bound $m_{\rm SUSY}=800\ {\rm GeV}$ is determined from the present 
discrepancy between values of the muon magnetic moment obtained from the 
$e^+e^-$ data and $\tau$ data \cite{g-2} 
which cannot be explained within the SM at the moment. 
Within so small range of $m_{\rm SUSY}$ values, it is hard to 
determine which of the amplitudes dominates without detailed numerical study. 
The numerical values show that the different processes have different 
behaviour, with softer $m_{\rm SUSY}$ for $\ell \to \ell'\gamma$ 
processes. That indicates that the contributions of some of three 
contributions compete for dominance in the branching ratio. Again, among all 
the processes the $\mu\to e\gamma$ is the most interesting 
from the experimental point of view. All numerical results 
for both model A and B are shown in Table 2. 
For model B we find that in the parameter region, 
$560\ \mbox{GeV} \leq m_{\rm SUSY} \leq 800\ \mbox{GeV}$ 
consistent with all the experimental data, 
the predicted values for the semileptonic LFV branching ratios are
close to the corresponding current experimental upper bounds. 
Therefore, they may be tested in the near future.

\begin{table*}[t]
\setcounter{table}{1}
\caption{The branching ratios for neutrinoless SL LFV of a charged lepton 
and and $\ell \to \ell' \gamma$ processes.}
\label{table:2}
\newcommand{\m}{\hphantom{$-$}}
\newcommand{\cc}[1]{\multicolumn{1}{c}{#1}}
\renewcommand{\tabcolsep}{1.9pc} 
\renewcommand{\arraystretch}{1.1} 
\begin{tabular}{@{}llll}\\
\hline\hline
decay ~modes &
exp.~bound &
model ~A &
model ~B \\
\hline\hline
$\mu\to e^-\gamma$                & $1.2\times 10^{-11}$ & 
$8.1\times 10^{-9} x^2_{\mu e}$   & $1.2\times 10^{-11}$ \\
$\tau\to e^-\gamma$               & $3.8\times 10^{-7}$ & 
$3.4\times 10^{-8} x^2_{\tau e}$  & 
$\left(2.6\times 10^{-10},~1.0\times 10^{-9} \right)$ \\
$\tau\to \mu^-\gamma$    & $3.1\times 10^{-7}$ & 
$6.7\times 10^{-9} x^2_{\tau\mu}$ & 
$\left(1.1\times 10^{-9}, ~4.5\times 10^{-9} \right)$ \\
\hline
$\tau\to e^-\pi^0$                & $1.9\times 10^{-7}$ & 
$2.8\times 10^{-6} y^2_{\tau e}$  & 
$\left(7.4\times 10^{-9},~4.1\times 10^{-8} \right)$ \\
$\tau\to \mu^-\pi^0$              & $4.3\times 10^{-7}$ & 
$5.4\times 10^{-7} y^2_{\tau\mu}$ & 
$\left(3.4\times 10^{-8},~1.8\times 10^{-7}\right)$ \\
$\tau\to e^-\eta$                 & $2.3\times 10^{-7}$ & 
$4.0\times 10^{-7} y^2_{\tau e}$  & 
$\left(1.1\times 10^{-9},~6.2\times 10^{-9} \right)$ \\
$\tau\to \mu^-\eta$               & $1.3\times 10^{-7}$ & 
$7.8\times 10^{-8} y^2_{\tau\mu}$ & 
$\left(5.1\times 10^{-9},~1.3\times 10^{-8} \right)$ \\
$\tau\to e^-\rho^0$               & $2.0\times 10^{-6}$ & 
$2.7\times 10^{-6} y^2_{\tau e}$  & 
$\left(4.0\times 10^{-14},~2.9\times 10^{-14}\right)$ \\
$\tau\to \mu^-\rho^0$             & $6.3\times 10^{-6}$ & 
$5.3\times 10^{-7} y^2_{\tau\mu}$ & 
$\left(1.3\times 10^{-13},~1.2\times 10^{-12} \right)$ \\
$\tau\to e^-\phi$                 & $6.9\times 10^{-6}$ & 
$2.7\times 10^{-6} y^2_{\tau e}$  & 
$\left(9.6\times 10^{-10},~5.1\times 10^{-9} \right)$ \\
$\tau\to \mu^-\phi$               & $7.0\times 10^{-6}$ & 
$5.3\times 10^{-7} y^2_{\tau\mu}$ & 
$\left(4.1\times 10^{-9},~2.2\times 10^{-8} \right)$ \\
\hline
$\tau\to e^-\pi^+\pi^-$  & $2.2\times 10^{-6}$ & 
$2.7\times 10^{-6} y^2_{\tau e}$ & \\
$\tau\to \mu^-\pi^+\pi^-$ & $8.2\times 10^{-6}$ & 
$5.2\times 10^{-7} y^2_{\tau\mu}$ & \\
$\tau\to e^-K^+K^-$ & $6.0\times 10^{-6}$ & 
$1.1\times 10^{-6} y^2_{\tau e}$ & \\
$\tau\to \mu^-K^+K^-$ & $1.5\times 10^{-5}$ & 
$2.1\times 10^{-7} y^2_{\tau\mu}$ & \\
\hline\hline
\end{tabular}\\[2pt]
The model A values are upper limits of branching ratios
multiplied by combinations of $B_{lN}$ 
matrix elements that are always smaller than one \cite{Amon00}. 
For model B the first(second) value in brackets represents lower(upper)
bound for a branching ratio. For $\mu \to e \gamma$ only the upper
bound is given, which served as an input parameter to determine the minimal $m_{SUSY}$ 
value.
\end{table*}

\section*{\bf Acknowledgments}

A.I. would like to thank to I. Picek and S. Fajfer for
the discussions on current-quark masses and evaluation of
tensor-quark current.
A part of the work was presented by A.I. at the workshop, 
``The 8th International Workshop on Tau-Lepton Physics (Tau04)'', 
held in Nara-ken New Public Hall, Japan. 
We are grateful to all organizers of this workshop and 
particularly to Prof. Ohshima for his kind hospitality 
extended to A.I and T.K. during their stay at Nagoya University. 
T.F. would like to thank S.T. Petcov for his hospitality
at SISSA.
This work of T.F. and T.K. was supported by 
the Grant in Aid for Scientific Research from the Ministry of 
Education, Science and Culture of Japan and the work of T.K. was 
supported by the Research Fellowship 
of the Japan Society for the Promotion of Science (\# 16540269 
and \# 7336). The work of A.I was supported by the Ministry of 
Science and Technology of Republic of Croatia under contract 
0119261.


\begin{thebibliography}{9}
\bibitem{EW86}
E. Witten, Nucl. Phys. B 268, 79 (1986). 
\bibitem{RMiV86}
R.N. Mohapatra and J.W.F. Valle, 
Phys. Rev. D 34, 1642 (1986). 
\bibitem{Val90}
M. Dittmar, A. Santamaria, M.C. Gonzalez-Garcia and J.W.F. Valle, 
Nucl. Phys. B 332, 1 (1990). 
\bibitem{Babu93}
K.S. Babu and R.N. Mohapatra,
Phys. Rev. Lett. 70, 2845 (1993).
\bibitem{FO02}
T. Fukuyama and N. Okada, JHEP 0211, 011 (2002);
K. Matsuda, Y. Koide, T. Fukuyama and H. Nishiura,
Phys. Rev. D 65, 033008 (2002), [Erratum-ibid. 
D 65, 079904 (2002)]; 
K. Matsuda, Y. Koide and T. Fukuyama, 
Phys. Rev. D 64, 053015 (2001).
\bibitem{Goh03}
H.S. Goh, R.N. Mohapatra and Siew-Phang Ng, 
Phys. Lett. B 570, 215 (2003); 
Phys. Rev. D 68, 115008 (2003).
\bibitem{Senj03}
B. Bajc, G. Senjanovi\'c and F. Vissani, 
Phys. Rev. Lett. 90, 051802 (2003).
\bibitem{Mim04}
B. Dutta, Y. Mimura and R.N. Mohapatra, 
Phys. Rev. D 69, 115014 (2004).
\bibitem{Mat04}
K. Matsuda, Phys. Rev. D 69, 113006 (2004). 
\bibitem{Masiero86}
F. Borzumati, A. Masiero, 
Phys. Rev. Lett. 57, 961 (1986).
\bibitem{Hisano96}
J. Hisano, T. Moroi, K. Tobe and M. Yamaguchi,
Phys. Rev. D 53, 2442 (1996).
\bibitem{mSUGRA}
R.~Barbieri, S.~Ferrara and C.A.~Savoy,
Phys. Lett. B 119, 343 (1982);
A.H.~Chamseddine, R.~Arnowitt and P.~Nath,
Phys. Rev. Lett. 49, 970 (1982);
L.J.~Hall, J.~Lykken and S.~Weinberg,
Phys. Rev. D 27, 2359 (1983). 
\bibitem{YE_tau04}
Y. Enari  [Belle collaboration], 
talk given at this workshop, 
[http://www.hepl.phys. nagoya-u.ac.jp/ public/Tau04/]. 
\bibitem{Val86}
J.W.F. Valle, in 
NUCLEAR BETA DECAYS AND NEUTRINO: proceedings. 
Edited by T. Kotani, H. Ejiri, E. Takasugi. 
Singapore, World Scientific, 1986. 542p. 

\bibitem{GGVal89}
M.C. Gonzalez-Garcia and J.W.F. Valle, Phys. Lett.
B 216 (1989).
\bibitem{BerVal87}
J. Bernab\'eu, A. Santamaria, J. Vidal, A. Mendez and J.W.F. Valle,
Phys. Lett. B 187, 303 (1997).
\bibitem{RiVal90}
N. Riuz and J.W.F. Valle,
Phys. Lett. B 246, 249 (1990).
\bibitem{GGVal92}
M.C. Gonzalez-Garcia and J.W.F. Valle,
Mod. Phys. Lett. A 7, 477 (1992).
\bibitem{AmApos95}
A. Ilakovac and A. Pilaftsis,
Nucl. Phys. B 437, 491 (1995).
\bibitem{sL88_92}
P. Langacker and D. London,
Phys. Rev. 38 (1988) 886;
E. Bhattachatyya, A. Datta, S.N. Ganguli and A. Raychaudhuri,
Mod. Phys. Lett A 6, 2991 (1991);
E. Nardi, E. Roulet and D. Tommasini,
Nucl. Phys. B 386, 239 (1992).
\bibitem{sLTomass95}
D. Tommasini, G. Barneboim, J. Bernab\'eu and C. Jarlskog,
Nucl. Phys. B 444, 451 (1995).
\bibitem{Tomass94}
E. Nardi, E. Roulet and D, Tommasini,
Phys. Lett. B 327, 319 (1994).
\bibitem{TATSNI04}
T. Fukuyama, A. Ilakovac, T. Kikuchi, S. Meljanac 
and N. Okada, hep-ph/0401213.
\bibitem{TATSNII04}
T. Fukuyama, A. Ilakovac, T. Kikuchi, S. Meljanac 
and N. Okada, to be published in J. Math. Phys. 
\bibitem{TAT04}
T. Fukuyama, A. Ilakovac, T. Kikuchi, 
in preparation. 
\bibitem{TATSNIII04}
T. Fukuyama, A. Ilakovac, T. Kikuchi, S. Meljanac 
and N. Okada, JHEP 0409 (2004) 052. [hep-ph/0401213]. 
\bibitem{Marshak69}
R.E. Marshak, Riazuddin and C.P. Ryan, 
{\it Weak Interactions in Particle Physics},
(Wiley, New York, 1969).
\bibitem{Sakurai69}
J.J. Sakurai, {\it Currents and Mesons}, 
(University of Chicago Press, Chicago, 1969).
\bibitem{Fubini73}
V. de Alfaro, S. Fubini and C. Rossetti, 
{\it Currents in Hadron Physics}, 
(North-Holland, Amsterdam, 1973).
\bibitem{Zielinski87}
M. Zielinski, Acta Phys. Polon. B 18, 455 (1987).
\bibitem{Gerard87}
W. A. Bardeen, A. J. Buras, and J.-M. G\'erard,
Nucl. Phys. B 293, 787 (1987); 
W. A. Bardeen, A. J. Buras, and J.-M. G\'erard,
Phys. Lett. B 180, 133 (1986);
R. S. Chivukula, J.M. Flynn and H. Georgy, 
{\it ibid.} 171, 453 (1986).
\bibitem{GassLeut84}
J. Gasser and H. Leutwyler, Ann. Phys. 158, 142 (1994).
\bibitem{Bando88}
M. Bando, T. Kugo, K. Yamawaki, 
Phys. Rep. 164, 217 (1988).
\bibitem{Bando85}
M. Bando, T. Kugo, K. Yamawaki, 
Nucl. Phys. B 259, 493 (1985).
\bibitem{FajAm98}
S. Fajfer and A. Ilakovac, Phys. Rev. D 57, 4219 (1998). 
\bibitem{Amon00}
A. Ilakovac, Phys. Rev. D 62, 036010 (2000).
\bibitem{g-2}
G.~W.~Bennett {\it et al.}  [Muon g-2 Collaboration],
 Phys.\ Rev.\ Lett.\  89, 101804 (2002)
 [Erratum-ibid.\  89, 129903 (2002)]
 [arXiv:hep-ex/0208001]. 
\end{thebibliography}
\end{document}